\documentclass[aip,
 amsmath,amssymb,
 reprint
]{revtex4-2}

\usepackage{graphicx}
\usepackage{dcolumn}
\usepackage{bm}
\usepackage{hyperref}
\hypersetup{colorlinks,linkcolor={blue},citecolor={blue},urlcolor={red}}
\usepackage[utf8]{inputenc}
\usepackage[T1]{fontenc}
\usepackage{mathptmx}
\usepackage{etoolbox}
\def\@email#1#2{%
 \endgroup
 \patchcmd{\titleblock@produce}
  {\frontmatter@RRAPformat}
  {\frontmatter@RRAPformat{\produce@RRAP{*#1\href{mailto:#2}{#2}}}\frontmatter@RRAPformat}
  {}{}
}%
\makeatother

\begin{document}


\title{High-fidelity QND readout and measurement back-action in a Tantalum-based high-coherence fluxonium qubit}

\author{Gaurav Bothara}
\author{Srijita Das}
\author{Kishor V Salunkhe}
\author{Madhavi Chand}
\author{Jay Deshmukh}
\author{Meghan P Patankar}
\author{R Vijay}
\email{r.vijay@tifr.res.in}
\affiliation{ 
Department of Condensed Matter Physics and Materials Science, Tata Institute of Fundamental Research, Homi Bhabha Road, Colaba, Mumbai-400005, India
}

\date{\today}

\begin{abstract}
Implementing a precise measurement of the quantum state of a qubit is very critical for building a practical quantum processor as it plays an important role in state initialization and quantum error correction. While the transmon qubit has been the most commonly used design in small to medium-scale processors, the fluxonium qubit is emerging as a strong alternative with the potential for high-fidelity gate operation as a result of the high anharmonicity and high coherence achievable due to its unique design. Here, we explore the measurement characteristics of a tantalum-based high-coherence fluxonium qubit and demonstrate single-shot measurement fidelity (assignment fidelity) of 96.2\%   and 97.8\%  without and with the use of a Josephson Parametric Amplifier respectively. We study the back-action of the measurement photons on the qubit and measure a QND (repeatability) fidelity of 99.6\%.  We find that the measurement fidelity and QND nature are limited by state-mixing errors and our results suggest that a careful study of measurement-induced transitions in the fluxonium is needed to further optimize the readout performance. 
\end{abstract}
\maketitle

\section {Introduction}
High fidelity, single-shot, Quantum Non-Demolition (QND) measurement is one of the essential tools in realizing some of the Di Vincenzo criteria for quantum information processing \cite{divincenzo_physical_2000}. QND readout perfectly correlates states post-readout with the subsequent measurement outcomes and guarantees the qubit is in computational subspace \cite{jeffrey_fast_2014,gusenkova_quantum_2021}. QNDness is crucial in realizing experimental protocols that require measurement-based qubit initialization\cite{houck_controlling_2008,sunada_fast_2022}. Moreover, fast QND readout is necessary for implementing cyclic quantum error correction (QEC) protocols\cite{bravyi_high-threshold_2024,fowler_surface_2012,chen_exponential_2021}, especially those with strict qubit initialization requirements on the measurement of syndrome qubit. QND character of the measurement is needed for surface codes\cite{fowler_surface_2012}, where each measurement cycle needs to be several orders faster than the qubit coherence times. 
\par In quantum information processing, superconducting qubit platforms are one of the forerunners. QND readout in superconducting qubits is typically achieved by transverse coupling between the qubit and the resonator\cite{jaynes_comparison_1963} where the resonator frequency is dependent on the qubit's state in the dispersive regime\cite{blais_cavity_2004,wallraff_strong_2004}. Speed and fidelity of single-shot measurement depend on several factors like dispersive shift $\chi$, photon decay rate $\kappa$, average readout photons $\Bar{n}$, integration time, and qubit coherence. A straightforward approach for faster and better readout is increasing the readout drive power, which increases the average photons $\Bar{n}$ in the resonator. However, the simple picture of dispersive readout fails when increasing readout power introduces non-QND effects such as measurement-induced state transitions (MIST) \cite{sank_measurement-induced_2016,nesterov_measurement-induced_2024}. 
\par Among superconducting qubits, transmon is the standard choice for constructing quantum processors\cite{koch_charge-insensitive_2007,arute_quantum_2019,zhang_high-performance_2022} due to the simplicity and robustness of the design. However, the transmon is a weakly anharmonic qubit, and anharmonicity is a precious resource in building densely connected scalable processors\cite{arute_quantum_2019,zhang_high-performance_2022}. In recent times, fluxonium has emerged as a serious challenger to the transmon, with its coherence times reaching milli-seconds and its highly anharmonic energy spectrum\cite{manucharyan_fluxonium_2009,nguyen_high-coherence_2019,somoroff_millisecond_2023}. While most of the recent work on fluxonium has focused on the characterization of losses\cite{vool_non-poissonian_2014,sun_characterization_2023} and the demonstration of single- and two-qubit gates\cite{ficheux_fast_2021,lin_24_2024,ma_native_2023,weiss_fast_2022,xiong_arbitrary_2022,dogan_two-fluxonium_2023}, the readout aspects of the fluxonium are relatively less explored. Recently, QND readout with high photon number ($\Bar{n}\sim200$) was demonstrated for fluxonium\cite{gusenkova_quantum_2021} with the use of granular aluminum (grAl) for the super-inductor.

\par In this article, we present a readout characterization of a tantalum-based fluxonium qubit\cite{nguyen_high-coherence_2019,somoroff_millisecond_2023} with a conventional super-inductor made using an array of overlap Josephson junctions and coherence times approaching the state-of-the-art. We achieved a single-shot measurement fidelity (assignment fidelity) of 96.2\% at a readout drive power with an average photon number $\Bar{n}\sim 112$ and an integration time of $2.82$ $\mu$s. Using Josephson parametric amplifiers (JPA), the best fidelity achieved is 97.8\% with $\Bar{n}\sim 126$ and integration time of $260$ ns and we also extract a QND fidelity if $99.6\%$. The fidelity is primarily limited by state preparation and mixing errors. We also explore resonator photons' back-action on the qubit state by performing a qubit decay experiment in the presence of resonator photons and see an indication of measurement-induced decay and MIST at high photon numbers.

\begin{table*}[ht]
\centering
\caption{\centering Device parameters (extracted from spectroscopy), and the measured frequencies and coherence times at half flux bias.}
		\begin{tabular}{c c c c c c c c c c c } 
			\hline
			\hline
			& Device Parameters &  & & Cavity & Cavity & Qubit & 1-2 transition &  &  Coherence &    \\
			$E_J/h$& $E_C/h$ & $E_L/h$ & &frequency & line width & frequency & frequency  & $T_{1,\mathrm{max}}$ & $T_{2,\mathrm{max}}^{R}$ & $T_{2,\mathrm{max}}$  \\
			 (GHz) &  (GHz) &  (GHz) & &${\omega_R}/{2\pi}$ (GHz) & ${\kappa_R}/{2\pi}$ (MHz) & ${\omega_{ge}}/{2\pi}$ (MHz) & ${\omega_{ef}}/{2\pi}$ (GHz)  & ($\mu$s) & ($\mu$s) & ($\mu$s) \\
			\hline 
			\hline
			\\
			4.098 & 0.754 & 0.998 & &7.167 & 11.6 & 328.12 & 3.062  & 402 & 298  & 627 \\ 
			\hline
		\end{tabular}
\label{table: device_parameter}
\end{table*}
\section{Device Characterization and Readout experiment}
Our fluxonium device has capacitor pads made of tantalum \cite{place2021new, wang2024achievingmillisecondcoherencefluxonium}, whereas the Josephson junctions are made of aluminum. The qubit frequency is $328.12$ MHz when biased at a magnetic field of half-flux quantum. The device parameters and coherence numbers are listed in Table \ref{table: device_parameter}. The fluxonium is transversely coupled to a 3D copper cavity with a readout frequency of $7.167$ GHz. The cavity has a weakly coupled port for qubit control and a strongly coupled port for measurement. The external coupling of the strong port is set to $\kappa_{R}/2\pi=11.6$ MHz and that of the weak port is set to $\approx \kappa_{R}/3$. Such a large $\kappa_R$ is chosen to reduce ambient photons in the cavity, effectively minimizing the dephasing caused by thermal photons\cite{nguyen_blueprint_2022}. Although the qubit is far detuned from the cavity, the higher fluxonium levels near the cavity frequency provide sufficient dispersive shift\cite{nguyen_blueprint_2022}. In our experiment, the cavity is placed between the $|g\rangle\rightarrow|h\rangle$ and $|e\rangle\rightarrow|i\rangle$ transitions where $|g\rangle$ and $|e\rangle$ are the ground and first excited states while $|h\rangle$ and $|i\rangle$ are the third and fourth excited states of the fluxonium qubit. We measure a total dispersive shift $ \chi_{ge} = 1.2$ MHz, at the half flux quanta bias point.
\begin{figure}[h]
    \centering
    \includegraphics[width=1.0\columnwidth]{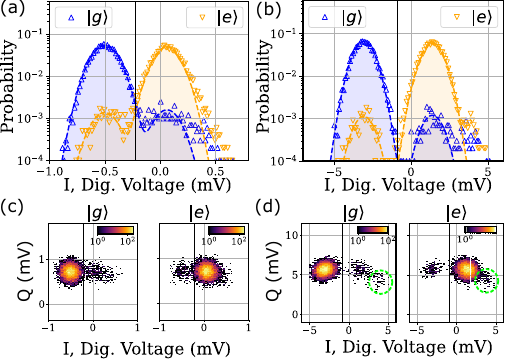} 
    \caption{Single-shot readout. Histograms with the double-gaussian fit for qubit prepared in $|g\rangle$ and $|e\rangle$ states at the optimal drive strength and integration time $\tau_{int}$ for readouts with JPA OFF (a) and ON (b).
(c) and (d) are IQ blobs at the mentioned optimal readout respectively for JPA OFF and ON.
The color axis represents counts. The green circular region in (d) indicates leakage out of the computational subspace. The black vertical solid lines are optimal binary discrimination thresholds.}
    \label{fig:opt_rdt}
\end{figure}
\par Since the fluxonium is a low-frequency qubit, it is expected to have a significant thermal population in the excited state when operating at typical dilution fridge temperatures. We measure the qubit in a mixed state with approximately 35\% of the population in the excited state corresponding to an effective qubit temperature of 25 mK. We initialize the qubit to the ground state using the two-photon sideband transition technique\cite{dogan_two-fluxonium_2023}. The excited-state population transitions to the readout resonator and since the resonator decays faster, it effectively cools the qubit. We achieve cooling of the qubit to about 5 mK (Details of the protocol are mentioned in the supplementary).
\subsection{Readout and QND fidelity}

 To measure the qubit, the readout cavity is driven via the strongly coupled port by deploying a square pulse at 7.167 GHz which is near the mean of the cavity frequencies corresponding to the qubit's ground and excited states. The reflected signal from the cavity is demodulated into the quadrature signals and integrated for time ($\tau_{int}$) giving a point in the IQ plane. On repeating single-shot measurements, we obtain a distribution of points in the IQ plane due to noise which we call an IQ blob from now on. To determine the single-shot fidelity of the readout, we prepare the qubit either in the ground state (no preparation pulse) or excited state ($\pi$ pulse) after the cooling sequence and then perform a single-shot measurement. This exercise is repeated 10,000 times to generate IQ blobs for each state. After this, an optimal threshold is chosen to discriminate between the measured states whose outcome is binary \{0,1\}, and the assignment fidelity is calculated as
 \begin{equation}
    \mathcal{F} = [P(0|g)+P(1|e)]/2
\end{equation}
 where $P(x|\psi)$ is the probability of measuring the outcome in $x$ given the state prepared is $|\psi\rangle$.  

\par To optimize readout fidelity, we adjust the readout drive's amplitude and the integration time of the measured output. For each amplitude, we alternate between the JPA pump being off and on and determine the assignment fidelity as a function of integration time. The best fidelity obtained without the JPA is 96.2\% which is achieved at $\Bar{n}\sim112$ readout photons in the cavity and $\tau_{int}=2.82 \mu$s. The histograms and IQ blobs for this operating point are shown in Fig.\ref{fig:opt_rdt}(a) and Fig.\ref{fig:opt_rdt}(c) respectively. The best readout fidelity with the JPA on is 97.8\% for a bias point with $\Bar{n}\sim126$ and $\tau_{int}=260 n$s and the corresponding data is shown in Fig.\ref{fig:opt_rdt}(b) and Fig.\ref{fig:opt_rdt}(d). The photon number in the cavity for a particular drive amplitude is calibrated by the Chi-Kappa-Power experiment\cite{sank_system_2024} (CKP, see Supplementary). The region in Fig.\ref{fig:opt_rdt}(d) highlighted by the green dashed circle indicates a transition rate to a state beyond the computational subspace of $\{|g\rangle,|e\rangle\}$ which becomes visible here due to the improved signal-to-noise ratio (SNR) with the JPA.

\begin{figure}[h]
    \centering
    \includegraphics[width=0.95\columnwidth]{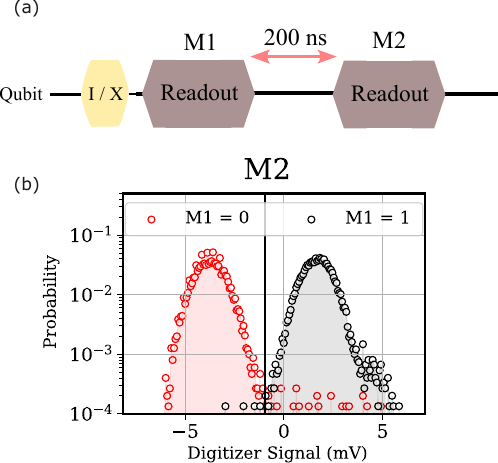} 
    \caption{ (a) QND experiment protocol. Measurement was done with JPA at the optimal readout amplitude and integration time. M1 and M2 are identical pulses with 340 ns length.  (b) Conditional histograms of M2 when M1 is measured to be either 0 or 1. The black solid vertical line is the threshold for optimal differentiation extracted by optimizing assignment fidelity for M1.}
    \label{fig:QND}
\end{figure} 
\begin{figure}[h]
    \centering
    \includegraphics[width=1.0\columnwidth]{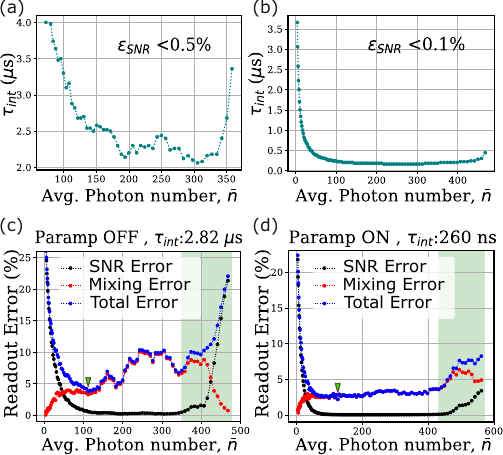} 
    \caption{Integration time versus readout photon number for $\varepsilon_{SNR}<0.5\%$ and $\varepsilon_{SNR}<0.1\%$ with the JPA (a) OFF and (b) ON. Readout errors as a function of readout photon number at optimal integration time $2.82$ $\mu$s and $260$ ns for parametric amplifier (c) OFF and (d) ON respectively. The green arrow indicates the optimal photon number. The green highlighted areas are the regions where $\varepsilon_{SNR}$ starts to increase unexpectedly.}
    \label{fig:err_rdt}
\end{figure}
\par Further, we test the QND nature of the measurement by performing two successive measurements (M1 and M2) with a 200 ns gap at the optimal readout parameters with JPA (pulse sequence shown in Fig.\ref{fig:QND}(a)). We use three state preparations with the qubit in state $|g\rangle$,  $|e\rangle$ or $(|g\rangle+|e\rangle)/2$ in consecutive experiments and implement both measurements for each case. The whole sequence is repeated $10000$ times and histograms of M2 when M1 is measured to be in 0 or 1 are shown in Fig.\ref{fig:QND}(b). The QND repeatability fidelity is given by $\mathcal{F}_Q$, where $\Bar{P}(a|b)$ is the probability of M2 being measured with an outcome b given that the M1 outcome was measured in a.
\begin{equation}
    \mathcal{F}_Q = [\Bar{P}(0|0)+\Bar{P}(1|1)]/2
\end{equation}
The demarcation value is chosen by optimizing assignment fidelity ($97.5\%$) for M1 and we use the same value to demarcate the histograms for M2 to assign the probabilities  $\Bar{P}(a|b)$. The calculated probabilities are $\Bar{P}(0|0)=0.995$ and $\Bar{P}(1|1)=0.997$. The repeatability fidelity measured is $\mathcal{F}_Q = 99.6\%$. As before, we see some signature of leakage beyond computational subspace in the right histogram (black circles) in Fig.\ref{fig:QND}(b) which corresponds to the M1 measurement projecting the qubit to the $|e\rangle$ state. Consequently, the $\mathcal{F}_Q$ slightly overestimates the true QND fidelity, as the protocol does not account for leakage outside the computational subspace for a certain fraction of the measurements\cite{hazra_benchmarking_2024}. These conditional histograms can also be interpreted in a different way. The M1 measurement can be considered as a way to herald a much purer state $|g\rangle$ (or $|e\rangle$) than the active cooling process. This is because of the high SNR and minimal overlap between the ground and excited state histograms. When intepreted this way, the calculated QND fidelity ($\mathcal{F}_Q$) for M2 is the same as the heraled assignment fidelity ($\mathcal{F}$) for M2 since $\Bar{P}(0|0)+ \Bar{P}(0|1)  =  \Bar{P}(1|0) + \Bar{P}(1|1) = 1$. The significant improvement in the heralded assignment fidelity is due to the near elimination of the state preparation error present in the previous measurement (Fig.\ref{fig:opt_rdt}(d)) due to ineffective cooling of the qubit.   

\begin{figure}[h]
    \centering
    \includegraphics[width=1.0\columnwidth]{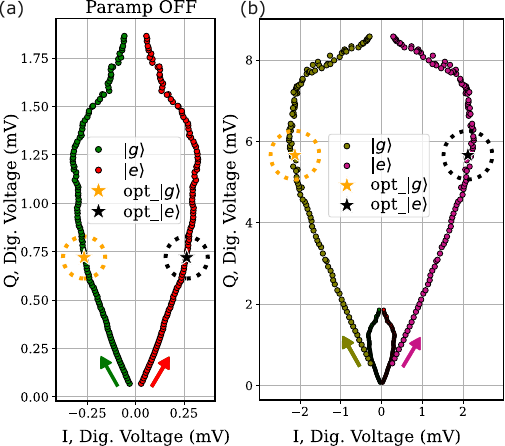} 
    \caption{Means of IQ blobs measured with the parametric amplifier (a) OFF and (b) both OFF and ON for qubit prepared in states $|g\rangle$ and $|e\rangle$. Data is shown different readout drive amplitude and the integration time corresponding to the best assignment fidelity at that amplitude. The arrows indicating increasing readout drive strength for respective states prepared. The stars indicate optimal readout amplitude for best assignment fidelity and the dashed circle indicates one standard deviation of the IQ distribution.}
    \label{fig:traj_rdt}
\end{figure}

\subsection{Readout Back-action}
The results so far demonstrate a high-fidelity readout of the fluxonium qubit but to understand the limiting factors and implement further improvements, we have to understand the different types of errors and their sources. In principle, increasing readout drive power should improve measurement fidelity by reducing the assignment error due to SNR or provide faster readout by achieving the same SNR at shorter integration times. However, increasing readout power may introduce state-mixing errors due to measurement-induced state transitions (MIST)\cite{sank_measurement-induced_2016,nesterov_measurement-induced_2024} within and beyond the computational subspace. To calculate the discrimination error due to SNR ($\varepsilon_{SNR}$), we fit a double Gaussian distribution to the readout histograms (as shown in Fig.\ref{fig:opt_rdt}(a,b)). The overlap area between the $|g\rangle$ and $|e\rangle$ state's dominant Gaussian is the $\varepsilon_{SNR}$ which is purely due to the SNR while the weights in the second Gaussian inform us about the error due to qubit population outside of the prepared state and can include state preparation and state mixing errors. 
\par Increasing readout power can enable us to do a faster readout as is exemplified by Fig.\ref{fig:err_rdt}(a,b), where we have shown the integration time needed to reduce  $\varepsilon_{SNR}$ below 0.005 and 0.001 with the JPA off and on respectively. For readout drives with photon number larger than $\sim108$ $\Bar{n}$, we achieve a measurement time-scale that is two orders of magnitude smaller than our coherence times (shown in Table.\ref{table: device_parameter}). Using the SNR numbers at low $\Bar{n}$, we extract the measurement efficiency without (with) the JPA as $\eta = 2.7\%$ (57.4\%) and the system noise temperature of our output chain as 12.9 K (0.6 K), which are typical for such cryogenic setups. As expected when using a JPA, we achieve high SNR and improved measurement fidelity with a measurement time reduced by one order of magnitude when compared to that without the JPA. To understand the effect of readout amplitude, we have plotted readout errors vs photon number in Fig.\ref{fig:err_rdt}(c,d) for $\tau_{int} = 2.82$ $\mu$s without the JPA and $\tau_{int} = 260$ ns with the JPA. We observe that while the error due to SNR improves with increasing readout power, the errors due to non-QND processes increase and limit the measurement fidelity beyond an optimal readout amplitude indicated by the green arrow in Fig.\ref{fig:err_rdt}(c,d).

\begin{figure}[h]
    \centering
    \includegraphics[width=1.0\columnwidth]{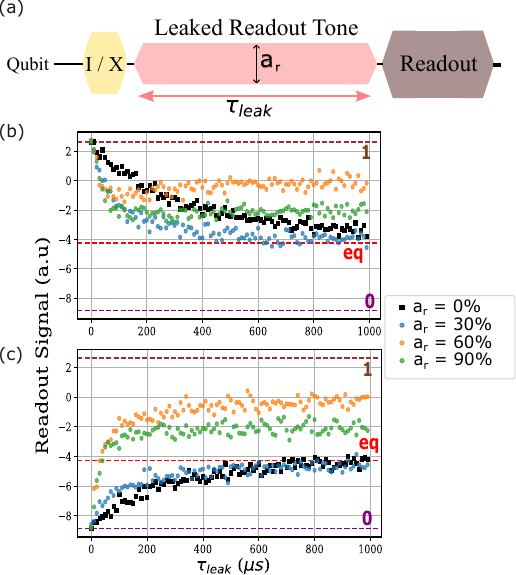} 
    \caption{(a) Pulse sequence used for state relaxation experiment in the presence of resonator photons. (b) and (c) show ensemble averaged signal vs $\tau_{leak}$ for states initialized in $|e\rangle$ and $|g\rangle$ respectively for different $a_r$, where $a_r$ is the fraction of amplitude used for optimal readout. Highlighted levels 0,1, and eq are averaged signals for $|g\rangle$,$|e\rangle$, and the thermal state of the qubit.  }
    \label{fig:leakage_exp}
\end{figure}

At even higher readout powers, the error due to SNR also increases unexpectedly as shown in the green highlighted regions in Fig.\ref{fig:err_rdt}(c,d)). We observed that the IQ blobs merge instead of separating at these higher readout powers which can be seen in Fig.\ref{fig:traj_rdt}(a,b) where we plot the mean of the IQ blobs for the qubit prepared in $|g\rangle$ and $|e\rangle$. The mean is plotted for $\tau_{int}$ corresponding to the optimal fidelity at that amplitude with and without the JPA. At lower powers, the means separate steadily as expected but start to merge after a certain readout power. This explains the unusual behavior of the $\varepsilon_{SNR}$ shown in the green highlighted region of Fig.\ref{fig:err_rdt}(c,d). We speculate that this may indicate that the qubit may be transitioning to states outside the computational subspace where the effective dispersive shift is smaller.

\par We further explore the readout photons' back-action on the qubit by performing a relaxation experiment in the presence of photons in the cavity and for much longer time scales. First, the qubit is prepared in either the $|g\rangle$ or $|e\rangle$ state after the cooling sequence. Then, a readout tone is introduced at a fraction ($a_r$) of the optimal readout amplitude for time $\tau_{leak}$ (pulse sequence shown in Figure \ref{fig:leakage_exp}(a)) and an ensemble measurement is performed using a second readout pulse with the optimal readout amplitude. The data is plotted or a few different fractional amplitudes $a_r$ in Figures \ref{fig:leakage_exp}(b) and \ref{fig:leakage_exp}(c) for the qubit prepared in $|e\rangle$ or $|g\rangle$ respectively. For small $a_r=30\%$ we observe the qubit decaying to the thermal equilibrium state. However, transition rates to equilibrium are significantly higher than in the case of no leaked photons ($a_r=0\%$) indicative of enhanced state mixing within the computational subspace. However, at higher values of $a_r$, the signal not only relaxes faster but also saturates to levels above the halfway mark between the \( |e\rangle \) or \( |g\rangle \) state signals. Such an observation is clear evidence of leakage outside the computational where a different value of dispersive shift causes the ensemble-averaged readout signal to show such behavior.

\section{Conclusion}
 In conclusion, we demonstrated high-fidelity QND readout in a tantalum-based fluxonium qubit with coherence times approaching the state-of-the-art. Despite the small $\chi_{ge}/\kappa_r$ ratio and without using a JPA, we achieve a single-shot measurement assignment fidelity of 96.2\% with an average photon number $\Bar{n}\sim112$ and an integration time of $2.82 \mu$s. This improves to 97.8\% when using a JPA with  $\Bar{n}\sim126$ and an integration time of $260$ ns. The improvement with the JPA is modest as measurement fidelity in both cases is dominated by state preparation and mixing errors and the errors due to SNR are negligible (0.5\% without JPA and 0.01\% with JPA).  We also measure a QND (repeatability) fidelity $\mathcal{F}_Q = 99.6\%$.  The readout fidelity at high drive powers drops significantly, indicating strong back-action due to the measurement photons. The back-action experiment qualitatively shows us accelerated decay rates and population leaking outside computational subspace similar to what has been observed in transmon qubits\cite{PhysRevApplied.20.054008}. These transitions beyond computational sub-space have been extensively investigated in transmon qubits and are referred to as qubit ionization\cite{PhysRevX.14.041023}. Our work elucidates that while a high-fidelity QND readout is possible with the fluxonium qubit, more theoretical\cite{nesterov_measurement-induced_2024,singh2024impactjosephsonjunctionarray} and experimental\cite{stefanski_improved_2024} investigations are needed to further understand and improve the readout and make progress towards a fluxonium based quantum processor. 

\section*{Supplementary material}
The supplementary material has three sections and provides more details: qubit reset; photon number calibration, CKP; and noise temperature and measurement efficiency. 
\section*{Acknowledgements}
We thank Mandar Deshmukh for his feedback on the paper. This work is supported by the Department of Atomic Energy of the Government of India under Project No. RTI4003. We also acknowledge funding from the National Mission on Interdisciplinary CyberPhysical Systems (NM-ICPS) of the Department of Science and Technology, Govt. of India through the I-HUB Quantum Technology Foundation. We acknowledge the TIFR Central Workshop and the Nanofabrication facility. 
\section*{Author Declarations}

\subsection*{Conflict of Interest}
The authors have no conflicts of interest to disclose.

\subsection*{Authors Contribution}
\noindent \textbf{Gaurav Bothara}: Conceptualization (equal); Formal analysis (lead); Methodology (equal); Software (equal); Writing -- original draft (lead); Writing -- review and editing (equal).
\textbf{Srijita Das}: Methodology (equal); Writing -- review and editing (supporting).
\textbf{Kishor V Salunkhe}: Software (equal); Conceptualization (supporting).
\textbf{Madhavi Chand}: Methodology (equal); Writing -- review and editing (supporting).
\textbf{Jay Deshmukh}: Software (equal); Writing -- review and editing (supporting).
\textbf{Meghan P Patankar}: Methodology (supporting); Supervision (supporting).
\textbf{R Vijay}: Conceptualization (equal); Supervision (lead); Writing -- original draft (supporting); Writing -- review and editing (equal).

\section*{References}
\bibliography{main}

\begin{thebibliography}{10}
\expandafter\ifx\csname url\endcsname\relax
  \def\url#1{\texttt{#1}}\fi
\expandafter\ifx\csname urlprefix\endcsname\relax\def\urlprefix{URL }\fi
\providecommand{\bibinfo}[2]{#2}
\providecommand{\eprint}[2][]{\url{#2}}

\bibitem{divincenzo_physical_2000}
\bibinfo{author}{DiVincenzo, D.~P.}
\newblock \bibinfo{title}{The {Physical} {Implementation} of {Quantum} {Computation}}.
\newblock \emph{\bibinfo{journal}{Fortschritte der Physik}} \textbf{\bibinfo{volume}{48}}, \bibinfo{pages}{771--783} (\bibinfo{year}{2000}).
\newblock \urlprefix\url{https://onlinelibrary.wiley.com/doi/abs/10.1002/1521-3978%28200009%2948%3A9/11%3C771%3A%3AAID-PROP771%3E3.0.CO%3B2-E}.
\newblock \bibinfo{note}{\_eprint: https://onlinelibrary.wiley.com/doi/pdf/10.1002/1521-3978\%28200009\%2948\%3A9/11\%3C771\%3A\%3AAID-PROP771\%3E3.0.CO\%3B2-E}.

\bibitem{jeffrey_fast_2014}
\bibinfo{author}{Jeffrey, E.} \emph{et~al.}
\newblock \bibinfo{title}{Fast {Accurate} {State} {Measurement} with {Superconducting} {Qubits}}.
\newblock \emph{\bibinfo{journal}{Physical Review Letters}} \textbf{\bibinfo{volume}{112}}, \bibinfo{pages}{190504} (\bibinfo{year}{2014}).
\newblock \urlprefix\url{https://link.aps.org/doi/10.1103/PhysRevLett.112.190504}.
\newblock \bibinfo{note}{Publisher: American Physical Society}.

\bibitem{gusenkova_quantum_2021}
\bibinfo{author}{Gusenkova, D.} \emph{et~al.}
\newblock \bibinfo{title}{Quantum {Nondemolition} {Dispersive} {Readout} of a {Superconducting} {Artificial} {Atom} {Using} {Large} {Photon} {Numbers}}.
\newblock \emph{\bibinfo{journal}{Physical Review Applied}} \textbf{\bibinfo{volume}{15}}, \bibinfo{pages}{064030} (\bibinfo{year}{2021}).
\newblock \urlprefix\url{https://link.aps.org/doi/10.1103/PhysRevApplied.15.064030}.
\newblock \bibinfo{note}{Publisher: American Physical Society}.

\bibitem{houck_controlling_2008}
\bibinfo{author}{Houck, A.~A.} \emph{et~al.}
\newblock \bibinfo{title}{Controlling the {Spontaneous} {Emission} of a {Superconducting} {Transmon} {Qubit}}.
\newblock \emph{\bibinfo{journal}{Physical Review Letters}} \textbf{\bibinfo{volume}{101}}, \bibinfo{pages}{080502} (\bibinfo{year}{2008}).
\newblock \urlprefix\url{https://link.aps.org/doi/10.1103/PhysRevLett.101.080502}.
\newblock \bibinfo{note}{Publisher: American Physical Society}.

\bibitem{sunada_fast_2022}
\bibinfo{author}{Sunada, Y.} \emph{et~al.}
\newblock \bibinfo{title}{Fast {Readout} and {Reset} of a {Superconducting} {Qubit} {Coupled} to a {Resonator} with an {Intrinsic} {Purcell} {Filter}}.
\newblock \emph{\bibinfo{journal}{Physical Review Applied}} \textbf{\bibinfo{volume}{17}}, \bibinfo{pages}{044016} (\bibinfo{year}{2022}).
\newblock \urlprefix\url{https://link.aps.org/doi/10.1103/PhysRevApplied.17.044016}.
\newblock \bibinfo{note}{Publisher: American Physical Society}.

\bibitem{bravyi_high-threshold_2024}
\bibinfo{author}{Bravyi, S.} \emph{et~al.}
\newblock \bibinfo{title}{High-threshold and low-overhead fault-tolerant quantum memory}.
\newblock \emph{\bibinfo{journal}{Nature}} \textbf{\bibinfo{volume}{627}}, \bibinfo{pages}{778--782} (\bibinfo{year}{2024}).
\newblock \urlprefix\url{https://www.nature.com/articles/s41586-024-07107-7}.
\newblock \bibinfo{note}{Publisher: Nature Publishing Group}.

\bibitem{fowler_surface_2012}
\bibinfo{author}{Fowler, A.~G.}, \bibinfo{author}{Mariantoni, M.}, \bibinfo{author}{Martinis, J.~M.} \& \bibinfo{author}{Cleland, A.~N.}
\newblock \bibinfo{title}{Surface codes: {Towards} practical large-scale quantum computation}.
\newblock \emph{\bibinfo{journal}{Physical Review A}} \textbf{\bibinfo{volume}{86}}, \bibinfo{pages}{032324} (\bibinfo{year}{2012}).
\newblock \urlprefix\url{https://link.aps.org/doi/10.1103/PhysRevA.86.032324}.
\newblock \bibinfo{note}{Publisher: American Physical Society}.

\bibitem{chen_exponential_2021}
\bibinfo{author}{Chen, Z.} \emph{et~al.}
\newblock \bibinfo{title}{Exponential suppression of bit or phase errors with cyclic error correction}.
\newblock \emph{\bibinfo{journal}{Nature}} \textbf{\bibinfo{volume}{595}}, \bibinfo{pages}{383--387} (\bibinfo{year}{2021}).
\newblock \urlprefix\url{https://www.nature.com/articles/s41586-021-03588-y}.
\newblock \bibinfo{note}{Publisher: Nature Publishing Group}.

\bibitem{jaynes_comparison_1963}
\bibinfo{author}{Jaynes, E.} \& \bibinfo{author}{Cummings, F.}
\newblock \bibinfo{title}{Comparison of quantum and semiclassical radiation theories with application to the beam maser}.
\newblock \emph{\bibinfo{journal}{Proceedings of the IEEE}} \textbf{\bibinfo{volume}{51}}, \bibinfo{pages}{89--109} (\bibinfo{year}{1963}).
\newblock \urlprefix\url{https://ieeexplore.ieee.org/document/1443594}.
\newblock \bibinfo{note}{Conference Name: Proceedings of the IEEE}.

\bibitem{blais_cavity_2004}
\bibinfo{author}{Blais, A.}, \bibinfo{author}{Huang, R.-S.}, \bibinfo{author}{Wallraff, A.}, \bibinfo{author}{Girvin, S.~M.} \& \bibinfo{author}{Schoelkopf, R.~J.}
\newblock \bibinfo{title}{Cavity quantum electrodynamics for superconducting electrical circuits: {An} architecture for quantum computation}.
\newblock \emph{\bibinfo{journal}{Physical Review A}} \textbf{\bibinfo{volume}{69}}, \bibinfo{pages}{062320} (\bibinfo{year}{2004}).
\newblock \urlprefix\url{https://link.aps.org/doi/10.1103/PhysRevA.69.062320}.
\newblock \bibinfo{note}{Publisher: American Physical Society}.

\bibitem{wallraff_strong_2004}
\bibinfo{author}{Wallraff, A.} \emph{et~al.}
\newblock \bibinfo{title}{Strong coupling of a single photon to a superconducting qubit using circuit quantum electrodynamics}.
\newblock \emph{\bibinfo{journal}{Nature}} \textbf{\bibinfo{volume}{431}}, \bibinfo{pages}{162--167} (\bibinfo{year}{2004}).
\newblock \urlprefix\url{https://www.nature.com/articles/nature02851}.
\newblock \bibinfo{note}{Publisher: Nature Publishing Group}.

\bibitem{sank_measurement-induced_2016}
\bibinfo{author}{Sank, D.} \emph{et~al.}
\newblock \bibinfo{title}{Measurement-{Induced} {State} {Transitions} in a {Superconducting} {Qubit}: {Beyond} the {Rotating} {Wave} {Approximation}}.
\newblock \emph{\bibinfo{journal}{Physical Review Letters}} \textbf{\bibinfo{volume}{117}}, \bibinfo{pages}{190503} (\bibinfo{year}{2016}).
\newblock \urlprefix\url{https://link.aps.org/doi/10.1103/PhysRevLett.117.190503}.
\newblock \bibinfo{note}{Publisher: American Physical Society}.

\bibitem{nesterov_measurement-induced_2024}
\bibinfo{author}{Nesterov, K.~N.} \& \bibinfo{author}{Pechenezhskiy, I.~V.}
\newblock \bibinfo{title}{Measurement-induced state transitions in dispersive qubit-readout schemes}.
\newblock \emph{\bibinfo{journal}{Phys. Rev. Appl.}} \textbf{\bibinfo{volume}{22}}, \bibinfo{pages}{064038} (\bibinfo{year}{2024}).
\newblock \urlprefix\url{https://link.aps.org/doi/10.1103/PhysRevApplied.22.064038}.

\bibitem{koch_charge-insensitive_2007}
\bibinfo{author}{Koch, J.} \emph{et~al.}
\newblock \bibinfo{title}{Charge-insensitive qubit design derived from the {Cooper} pair box}.
\newblock \emph{\bibinfo{journal}{Physical Review A}} \textbf{\bibinfo{volume}{76}}, \bibinfo{pages}{042319} (\bibinfo{year}{2007}).
\newblock \urlprefix\url{https://link.aps.org/doi/10.1103/PhysRevA.76.042319}.
\newblock \bibinfo{note}{Publisher: American Physical Society}.

\bibitem{arute_quantum_2019}
\bibinfo{author}{Arute, F.} \emph{et~al.}
\newblock \bibinfo{title}{Quantum supremacy using a programmable superconducting processor}.
\newblock \emph{\bibinfo{journal}{Nature}} \textbf{\bibinfo{volume}{574}}, \bibinfo{pages}{505--510} (\bibinfo{year}{2019}).
\newblock \urlprefix\url{https://www.nature.com/articles/s41586-019-1666-5}.
\newblock \bibinfo{note}{Publisher: Nature Publishing Group}.

\bibitem{zhang_high-performance_2022}
\bibinfo{author}{Zhang, E.~J.} \emph{et~al.}
\newblock \bibinfo{title}{High-performance superconducting quantum processors via laser annealing of transmon qubits}.
\newblock \emph{\bibinfo{journal}{Science Advances}} \textbf{\bibinfo{volume}{8}}, \bibinfo{pages}{eabi6690} (\bibinfo{year}{2022}).
\newblock \urlprefix\url{https://www.science.org/doi/10.1126/sciadv.abi6690}.
\newblock \bibinfo{note}{Publisher: American Association for the Advancement of Science}.

\bibitem{manucharyan_fluxonium_2009}
\bibinfo{author}{Manucharyan, V.~E.}, \bibinfo{author}{Koch, J.}, \bibinfo{author}{Glazman, L.~I.} \& \bibinfo{author}{Devoret, M.~H.}
\newblock \bibinfo{title}{Fluxonium: {Single} {Cooper}-{Pair} {Circuit} {Free} of {Charge} {Offsets}}.
\newblock \emph{\bibinfo{journal}{Science}} \textbf{\bibinfo{volume}{326}}, \bibinfo{pages}{113--116} (\bibinfo{year}{2009}).
\newblock \urlprefix\url{https://www.science.org/doi/10.1126/science.1175552}.
\newblock \bibinfo{note}{Publisher: American Association for the Advancement of Science}.

\bibitem{nguyen_high-coherence_2019}
\bibinfo{author}{Nguyen, L.~B.} \emph{et~al.}
\newblock \bibinfo{title}{High-{Coherence} {Fluxonium} {Qubit}}.
\newblock \emph{\bibinfo{journal}{Physical Review X}} \textbf{\bibinfo{volume}{9}}, \bibinfo{pages}{041041} (\bibinfo{year}{2019}).
\newblock \urlprefix\url{https://link.aps.org/doi/10.1103/PhysRevX.9.041041}.
\newblock \bibinfo{note}{Publisher: American Physical Society}.

\bibitem{somoroff_millisecond_2023}
\bibinfo{author}{Somoroff, A.} \emph{et~al.}
\newblock \bibinfo{title}{Millisecond {Coherence} in a {Superconducting} {Qubit}}.
\newblock \emph{\bibinfo{journal}{Physical Review Letters}} \textbf{\bibinfo{volume}{130}}, \bibinfo{pages}{267001} (\bibinfo{year}{2023}).
\newblock \urlprefix\url{https://link.aps.org/doi/10.1103/PhysRevLett.130.267001}.
\newblock \bibinfo{note}{Publisher: American Physical Society}.

\bibitem{vool_non-poissonian_2014}
\bibinfo{author}{Vool, U.} \emph{et~al.}
\newblock \bibinfo{title}{Non-{Poissonian} {Quantum} {Jumps} of a {Fluxonium} {Qubit} due to {Quasiparticle} {Excitations}}.
\newblock \emph{\bibinfo{journal}{Physical Review Letters}} \textbf{\bibinfo{volume}{113}}, \bibinfo{pages}{247001} (\bibinfo{year}{2014}).
\newblock \urlprefix\url{https://link.aps.org/doi/10.1103/PhysRevLett.113.247001}.
\newblock \bibinfo{note}{Publisher: American Physical Society}.

\bibitem{sun_characterization_2023}
\bibinfo{author}{Sun, H.} \emph{et~al.}
\newblock \bibinfo{title}{Characterization of {Loss} {Mechanisms} in a {Fluxonium} {Qubit}}.
\newblock \emph{\bibinfo{journal}{Physical Review Applied}} \textbf{\bibinfo{volume}{20}}, \bibinfo{pages}{034016} (\bibinfo{year}{2023}).
\newblock \urlprefix\url{https://link.aps.org/doi/10.1103/PhysRevApplied.20.034016}.
\newblock \bibinfo{note}{Publisher: American Physical Society}.

\bibitem{ficheux_fast_2021}
\bibinfo{author}{Ficheux, Q.} \emph{et~al.}
\newblock \bibinfo{title}{Fast {Logic} with {Slow} {Qubits}: {Microwave}-{Activated} {Controlled}-{Z} {Gate} on {Low}-{Frequency} {Fluxoniums}}.
\newblock \emph{\bibinfo{journal}{Physical Review X}} \textbf{\bibinfo{volume}{11}}, \bibinfo{pages}{021026} (\bibinfo{year}{2021}).
\newblock \urlprefix\url{https://link.aps.org/doi/10.1103/PhysRevX.11.021026}.
\newblock \bibinfo{note}{Publisher: American Physical Society}.

\bibitem{lin_24_2024}
\bibinfo{author}{Lin, W.-J.} \emph{et~al.}
\newblock \bibinfo{title}{24 days-stable {CNOT}-gate on fluxonium qubits with over 99.9\% fidelity} (\bibinfo{year}{2024}).
\newblock \urlprefix\url{https://arxiv.org/abs/2407.15783v1}.

\bibitem{ma_native_2023}
\bibinfo{author}{Ma, X.} \emph{et~al.}
\newblock \bibinfo{title}{Native approach to controlled-$z$ gates in inductively coupled fluxonium qubits}.
\newblock \emph{\bibinfo{journal}{Phys. Rev. Lett.}} \textbf{\bibinfo{volume}{132}}, \bibinfo{pages}{060602} (\bibinfo{year}{2024}).
\newblock \urlprefix\url{https://link.aps.org/doi/10.1103/PhysRevLett.132.060602}.

\bibitem{weiss_fast_2022}
\bibinfo{author}{Weiss, D.} \emph{et~al.}
\newblock \bibinfo{title}{Fast {High}-{Fidelity} {Gates} for {Galvanically}-{Coupled} {Fluxonium} {Qubits} {Using} {Strong} {Flux} {Modulation}}.
\newblock \emph{\bibinfo{journal}{PRX Quantum}} \textbf{\bibinfo{volume}{3}}, \bibinfo{pages}{040336} (\bibinfo{year}{2022}).
\newblock \urlprefix\url{https://link.aps.org/doi/10.1103/PRXQuantum.3.040336}.
\newblock \bibinfo{note}{Publisher: American Physical Society}.

\bibitem{xiong_arbitrary_2022}
\bibinfo{author}{Xiong, H.} \emph{et~al.}
\newblock \bibinfo{title}{Arbitrary controlled-phase gate on fluxonium qubits using differential ac {Stark} shifts}.
\newblock \emph{\bibinfo{journal}{Physical Review Research}} \textbf{\bibinfo{volume}{4}}, \bibinfo{pages}{023040} (\bibinfo{year}{2022}).
\newblock \urlprefix\url{https://link.aps.org/doi/10.1103/PhysRevResearch.4.023040}.
\newblock \bibinfo{note}{Publisher: American Physical Society}.

\bibitem{dogan_two-fluxonium_2023}
\bibinfo{author}{Dogan, E.} \emph{et~al.}
\newblock \bibinfo{title}{Two-{Fluxonium} {Cross}-{Resonance} {Gate}}.
\newblock \emph{\bibinfo{journal}{Physical Review Applied}} \textbf{\bibinfo{volume}{20}}, \bibinfo{pages}{024011} (\bibinfo{year}{2023}).
\newblock \urlprefix\url{https://link.aps.org/doi/10.1103/PhysRevApplied.20.024011}.
\newblock \bibinfo{note}{Publisher: American Physical Society}.

\bibitem{place2021new}
\bibinfo{author}{Place, A.~P.} \emph{et~al.}
\newblock \bibinfo{title}{New material platform for superconducting transmon qubits with coherence times exceeding 0.3 milliseconds}.
\newblock \emph{\bibinfo{journal}{Nature communications}} \textbf{\bibinfo{volume}{12}}, \bibinfo{pages}{1779} (\bibinfo{year}{2021}).

\bibitem{wang2024achievingmillisecondcoherencefluxonium}
\bibinfo{author}{Wang, F.} \emph{et~al.}
\newblock \bibinfo{title}{Achieving millisecond coherence fluxonium through overlap josephson junctions} (\bibinfo{year}{2024}).
\newblock \urlprefix\url{https://arxiv.org/abs/2405.05481}.
\newblock \eprint{2405.05481}.

\bibitem{nguyen_blueprint_2022}
\bibinfo{author}{Nguyen, L.~B.} \emph{et~al.}
\newblock \bibinfo{title}{Blueprint for a {High}-{Performance} {Fluxonium} {Quantum} {Processor}}.
\newblock \emph{\bibinfo{journal}{PRX Quantum}} \textbf{\bibinfo{volume}{3}}, \bibinfo{pages}{037001} (\bibinfo{year}{2022}).
\newblock \urlprefix\url{https://link.aps.org/doi/10.1103/PRXQuantum.3.037001}.
\newblock \bibinfo{note}{Publisher: American Physical Society}.

\bibitem{sank_system_2024}
\bibinfo{author}{Sank, D.} \emph{et~al.}
\newblock \bibinfo{title}{System {Characterization} of {Dispersive} {Readout} in {Superconducting} {Qubits}} (\bibinfo{year}{2024}).
\newblock \urlprefix\url{http://arxiv.org/abs/2402.00413}.
\newblock \bibinfo{note}{ArXiv:2402.00413}.

\bibitem{hazra_benchmarking_2024}
\bibinfo{author}{Hazra, S.} \emph{et~al.}
\newblock \bibinfo{title}{Benchmarking the readout of a superconducting qubit for repeated measurements} (\bibinfo{year}{2024}).
\newblock \urlprefix\url{http://arxiv.org/abs/2407.10934}.
\newblock \bibinfo{note}{ArXiv:2407.10934 [physics, physics:quant-ph]}.

\bibitem{PhysRevApplied.20.054008}
\bibinfo{author}{Khezri, M.} \emph{et~al.}
\newblock \bibinfo{title}{Measurement-induced state transitions in a superconducting qubit: Within the rotating-wave approximation}.
\newblock \emph{\bibinfo{journal}{Phys. Rev. Appl.}} \textbf{\bibinfo{volume}{20}}, \bibinfo{pages}{054008} (\bibinfo{year}{2023}).
\newblock \urlprefix\url{https://link.aps.org/doi/10.1103/PhysRevApplied.20.054008}.

\bibitem{PhysRevX.14.041023}
\bibinfo{author}{Dumas, M.~F.} \emph{et~al.}
\newblock \bibinfo{title}{Measurement-induced transmon ionization}.
\newblock \emph{\bibinfo{journal}{Phys. Rev. X}} \textbf{\bibinfo{volume}{14}}, \bibinfo{pages}{041023} (\bibinfo{year}{2024}).
\newblock \urlprefix\url{https://link.aps.org/doi/10.1103/PhysRevX.14.041023}.

\bibitem{singh2024impactjosephsonjunctionarray}
\bibinfo{author}{Singh, S.}, \bibinfo{author}{Refael, G.}, \bibinfo{author}{Clerk, A.} \& \bibinfo{author}{Rosenfeld, E.}
\newblock \bibinfo{title}{Impact of josephson junction array modes on fluxonium readout} (\bibinfo{year}{2024}).
\newblock \urlprefix\url{https://arxiv.org/abs/2412.14788}.
\newblock \eprint{2412.14788}.

\bibitem{stefanski_improved_2024}
\bibinfo{author}{Stefanski, T.~V.} \emph{et~al.}
\newblock \bibinfo{title}{Improved fluxonium readout through dynamic flux pulsing} (\bibinfo{year}{2024}).
\newblock \urlprefix\url{http://arxiv.org/abs/2411.13437}.
\newblock \bibinfo{note}{ArXiv:2411.13437 [quant-ph]}.

\end{thebibliography}


\begin{thebibliography}{1}
\expandafter\ifx\csname url\endcsname\relax
  \def\url#1{\texttt{#1}}\fi
\expandafter\ifx\csname urlprefix\endcsname\relax\def\urlprefix{URL }\fi
\providecommand{\bibinfo}[2]{#2}
\providecommand{\eprint}[2][]{\url{#2}}

\bibitem{dogan_two-fluxonium_2023}
\bibinfo{author}{Dogan, E.} \emph{et~al.}
\newblock \bibinfo{title}{Two-{Fluxonium} {Cross}-{Resonance} {Gate}}.
\newblock \emph{\bibinfo{journal}{Physical Review Applied}} \textbf{\bibinfo{volume}{20}}, \bibinfo{pages}{024011} (\bibinfo{year}{2023}).
\newblock \urlprefix\url{https://link.aps.org/doi/10.1103/PhysRevApplied.20.024011}.
\newblock \bibinfo{note}{Publisher: American Physical Society}.

\bibitem{huber_parametric_2024}
\bibinfo{author}{Huber, G. B.~P.} \emph{et~al.}
\newblock \bibinfo{title}{Parametric multi-element coupling architecture for coherent and dissipative control of superconducting qubits} (\bibinfo{year}{2024}).
\newblock \urlprefix\url{http://arxiv.org/abs/2403.02203}.
\newblock \bibinfo{note}{ArXiv:2403.02203}.

\bibitem{zhou_rapid_2021}
\bibinfo{author}{Zhou, Y.} \emph{et~al.}
\newblock \bibinfo{title}{Rapid and unconditional parametric reset protocol for tunable superconducting qubits}.
\newblock \emph{\bibinfo{journal}{Nature Communications}} \textbf{\bibinfo{volume}{12}}, \bibinfo{pages}{5924} (\bibinfo{year}{2021}).
\newblock \urlprefix\url{https://www.nature.com/articles/s41467-021-26205-y}.
\newblock \bibinfo{note}{Publisher: Nature Publishing Group}.

\bibitem{sank_system_2024}
\bibinfo{author}{Sank, D.} \emph{et~al.}
\newblock \bibinfo{title}{System {Characterization} of {Dispersive} {Readout} in {Superconducting} {Qubits}} (\bibinfo{year}{2024}).
\newblock \urlprefix\url{http://arxiv.org/abs/2402.00413}.
\newblock \bibinfo{note}{ArXiv:2402.00413}.

\bibitem{hatridge_quantum_2013}
\bibinfo{author}{Hatridge, M.} \emph{et~al.}
\newblock \bibinfo{title}{Quantum {Back}-{Action} of an {Individual} {Variable}-{Strength} {Measurement}}.
\newblock \emph{\bibinfo{journal}{Science}} \textbf{\bibinfo{volume}{339}}, \bibinfo{pages}{178--181} (\bibinfo{year}{2013}).
\newblock \urlprefix\url{https://www.science.org/doi/10.1126/science.1226897}.
\newblock \bibinfo{note}{Publisher: American Association for the Advancement of Science}.

\end{thebibliography}
\end{document}


\title{Supplementary for "High-fidelity QND readout and measurement back-action in Tantalum-based high-coherence fluxonium qubit"}
\author{Gaurav Bothara}
\author{Srijita Das}
\author{Kishor V Salunkhe}
\author{Madhavi Chand}
\author{Jay Deshmukh}
\author{Meghan P Patankar}
\author{R Vijay}
\email{r.vijay@tifr.res.in}
\affiliation{Department of Condensed Matter Physics and Materials Science, Tata Institute of Fundamental Research, Homi Bhabha Road, Colaba, Mumbai-400005, India.}

\maketitle
\section {Qubit Reset}
Our fluxonium qubit equilibrium state has approximately 35\% population in the excited state. To "cool" the qubit we drive the qubit with a two-photon sideband transition from state $|e\rangle_{Q}|0\rangle_{R}$ and $|g\rangle_{Q}|1\rangle_{R}$\cite{dogan_two-fluxonium_2023,huber_parametric_2024,zhou_rapid_2021}, where we have assumed a system of qubit Q and resonator R. The excited state of the resonator  $|g\rangle_{Q}|1\rangle_{R}$ is short-lived and rapidly decays to $|g\rangle_{Q}|0\rangle_{R}$. 
\begin{figure}[h]
    \centering
    \includegraphics[width=0.95\columnwidth]{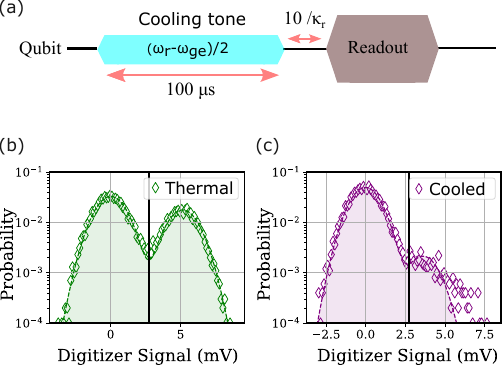} 
    \caption{ (a) Pulse sequence for the qubit cooling protocol. (b) and (c) Single-shot histograms without and with the cooling protocol implemented.}
    \label{fig:Cool}
\end{figure}

\par Without cooling, the qubit has an effective temperature of 25 mK calculated from the single-shot measurement shown in Fig.\ref{fig:Cool}(b). The pulse sequence for the cooling protocol is shown in Fig. \ref{fig:Cool}(a). The frequency of the cooling drive is set to $(f_R-f_Q)/2 = 3.42$ GHz. We then calibrate the drive power and the pulse duration of the cooling tone to maximize the reset to the ground state. Using the calibrated cooling tone, we minimized the residual excited population to $\approx 3\%$ as represented by the histogram in Fig.\ref{fig:Cool}(c).

\section {Photon Number Calibration, CKP}
\begin{figure}[h!]
    \centering
    \includegraphics[width=0.95\columnwidth]{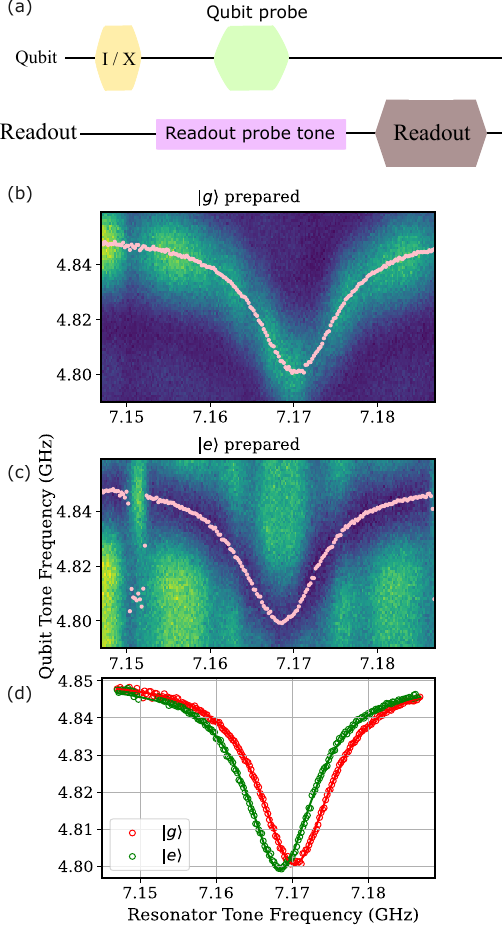} 
    \caption{CKP calibration for readout drive with an IF amplitude of 20 mV. (a) Pulse sequence for the calibration protocol. (b) and (c) 2D color plot for CKP measurements with qubit state prepared in $|g\rangle$ and $|e\rangle$ respectively with the color indicating the average readout signal. The pink dots correspond to the ac Stark-shifted qubit frequencies. (d) Lorentzian curve fit for the pink dots in (b) and (c).}
    \label{fig:CKP}
\end{figure}

We calibrate the readout drive power generated at room temperature to the photon number occupation in the cavity using the "chi-kappa-power" (CKP) experiment\cite{sank_system_2024}. We have performed this calibration at integer flux bias with the qubit frequency at 4.85 GHz. The pulse sequence for the CKP experiment is shown in Fig.\ref{fig:CKP}(a). The experiment measures the ac Stark shifted frequency of the qubit in the presence of photons on the resonator.
In the dispersive regime, when the resonator is populate with $\Bar{n}$ photons, it ac Stark shifts the qubit frequency by $\delta\omega_Q \approx \chi_{ge}\Bar{n}$. We first prepare the qubit in either $|g\rangle$ or $|e\rangle $ state and then drive the resonator at frequency $\omega_{d,R}$ long enough to populate the cavity steadily. Meanwhile, we drive the qubit at frequency $\omega_{d,Q}$ expecting the qubit to flip when the frequency matches the ac Stark-shifted frequency. Finally, we wait for the cavity to ring down and then measure the qubit's state.

\par We sweep the frequency $\omega_{d,R}$ and $\omega_{d,Q}$, and record the Stark-shifted qubit frequency as shown in Fig.\ref{fig:CKP}(b,c) which is a Lorentzian function of the readout drive frequency $\omega_{d,R}$. We fit the Lorentzian curves for each qubit state prepared as shown in Fig.\ref{fig:CKP}(d). We see some spurious data for the case of the excited state (see left side of Fig.\ref{fig:CKP}(c)) which has been omitted for the fits. Each of the fits gives a resonant frequency of the cavity dependent on the qubit state prepared, the difference in the resonant frequency gives a total dispersive shift $\chi_{ge}$. From the ac Stark-shift of the qubit at the cavity resonant frequency $\delta\omega_Q \approx \chi_{ge}\Bar{n}$, we calculate the photon occupation number $\Bar{n}$ for the incident readout power. Further, we assume the incident power at the cavity is proportional to power generated at room temperature. The data shown in Fig.\ref{fig:CKP} is taken for the 20 mV IF pulse generated for the readout drive. The generated readout drive power translates to $\Bar{n}=27\pm1$ photon occupation.

\section{Noise Temperature and Measurement Efficiency}

\begin{figure}[h]
    \centering
    \includegraphics[width=1\columnwidth]{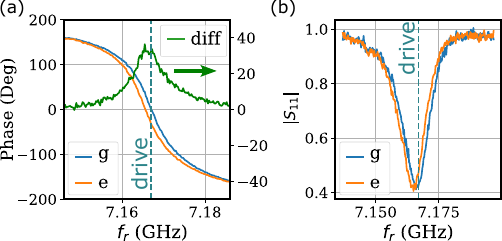} 
    \caption{ (a) Phase response of the resonator for qubit states $|g\rangle$ and $|e\rangle$. The phase difference is also plotted (right axis), and at $f_r=7.167$ GHz the difference is $\approx 32^{\circ}$. (b) The magnitude of the reflected signal from the resonator. The dashed vertical line represents the drive frequency.}
    \label{fig:cav_res}
\end{figure}
\begin{figure}[h]
    \centering
    \includegraphics[width=0.98\columnwidth]{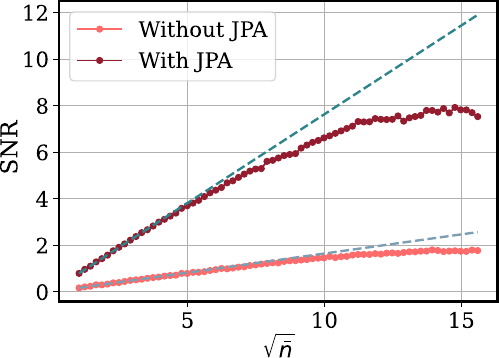} 
    \caption{ SNR calculated from the overlap of single-shot readout histograms for various readout power and plotted as a function of $\sqrt{\bar{n}}$ and $\tau_{int} = 1us$. The data is taken with and without JPA. The dashed line is a linear fit at a low photon number and is used to extract measurement efficiency and system noise temperature.}
    \label{fig:SNR}
\end{figure}

The signal-to-noise ratio\cite{hatridge_quantum_2013} for the qubit state measurement where information about the state is in I quadrature is given by, 
\begin{equation}
    \begin{aligned}
        SNR &= \frac{|\langle I_g\rangle - \langle I_e\rangle|}{\sigma_g+\sigma_e} \\
        &= \frac{\sqrt{n_m}}{\sqrt{n_n /2}} \; sin(\varphi) \\
    \end{aligned}
    \label{Eqn:SNR}
\end{equation}
where $\sigma_{g/e}$ is the variance of the output state histogram. Here $n_m$ is the measurement photon number, whereas $n_n$ is the added noise photon number. The $|g\rangle$ and $|e\rangle$ pointer states are separated by $2\varphi$ angle. For a measurement time $\tau_{int}$, $n_m=\Bar{n} \kappa \tau_{int} \times f$\cite{hatridge_quantum_2013}. In the case of a single-port driven loss-less resonator, the factor $f$ is $1/4$\cite{hatridge_quantum_2013}. For our device, the resonator has two ports with not-so-different external coupling. We measure the qubit by reflecting the readout pulse from the strong port with external coupling $\kappa_s = 11.6$ MHz. The strong port sees the other port as a lossy channel. The total linewidth of the cavity response reflected from the strong port is $\kappa = 15.6$ MHz, and the apparent internal loss is $\kappa^{\prime} = 4$ MHz.
\par The factor $f$ can be thought of as a ratio of powers, 
\begin{equation}
    f = \frac{P_{meas}}{P_{rad}}
\end{equation}
where $P_{rad}$ is the power radiated by a cavity with line-width $\kappa$ populated by $\Bar{n}$ photons. $P_{meas}$ is readout signal power reflected from driving the strong port with cavity population maintained at $\Bar{n}$.In addition to the 1/4 factor, the two-port resonator driven from a strong port requires a bit more power to maintain the same $\Bar{n}$ (two-port resonator is simulated in `AWR Microwave office'). Furthermore, a correction for the reflection coefficient is necessary. We take the mean of the reflection response (shown in Fig.\ref{fig:cav_res}(b)) at the readout drive. The power ratio $f$ is $-11.67$ dB. The angle $2\varphi$ is taken as the phase difference at the readout drive frequency (see Fig.\ref{fig:cav_res}(a)). We substitute $f$ and $\varphi$, and fit a straight line to the SNR as a function of $\sqrt{\Bar{n}}$ (shown in Fig.\ref{fig:SNR}). 

\par
\begin{figure}[t]
    \centering
    \includegraphics[width=1.0\columnwidth]{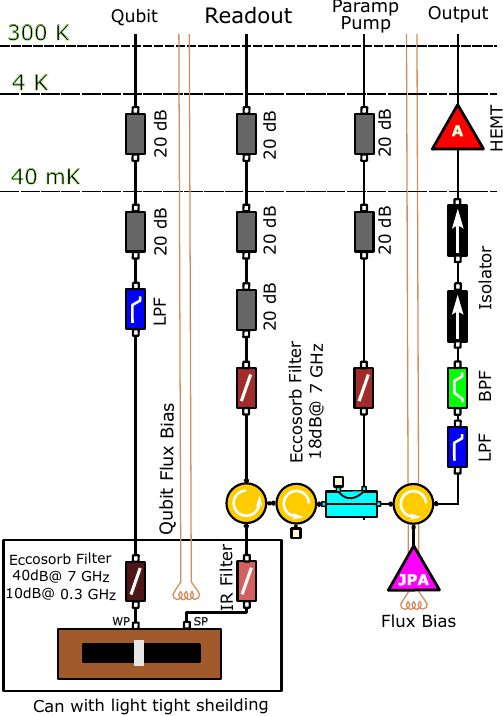} 
    \caption{ Experimental setup and microwave components at various stages of the dilution refrigerator. The bottom half of the copper cavity is shown with a qubit chip at the center.}
    \label{fig:supp_wire}
\end{figure}

 The calculated added noise photon number is $n_n = 37.5$ for the measurement without the Josephson parametric amplifier (JPA). The corresponding effective noise temperature is $T_{n,eff}= n_n \hbar \omega_r/k_B \approx 12.9$ K. This number is compatible with the typical noise temperature of the HEMT amplifiers and the expected insertion loss in the output line between the device and the HEMT due to various circulators/isolators and cables. With JPA, the added photon noise is  $n_n=1.7$ and $T_{n,eff}\approx 0.6$ K. The measurement efficiency is defined as $\eta = 2\sigma^2_0/n_n$, where $\sigma_0= 1/\sqrt{2}$ is the quantum-limited variance for the measurement output. The measurement efficiency achieved for the measurement chain shown in Fig.\ref{fig:supp_wire} is $\eta= 2.7\%$ without JPA and $\eta= 57.4\%$ with JPA.
 
\section*{References}
\bibliography{supp}